\documentclass[journal,10pt]{IEEEtran}

\usepackage{amsmath}
\usepackage{suffix}
\usepackage{mathtools}
\usepackage{cite}
\usepackage{color,soul}
 
\usepackage{multirow}
\usepackage{hhline}

\ifCLASSINFOpdf
\else
\fi

\usepackage{suffix,mathtools}
\DeclarePairedDelimiterX\MeijerM[3]{\lparen}{\rparen}%
{#3\,\delimsize\vert\begin{smallmatrix}#1 \\ #2\end{smallmatrix}}

\newcommand\MeijerG[8][]{%
  G^{\,#2,#3}_{#4,#5}\MeijerM[#1]{#6}{#7}{#8}}

\WithSuffix\newcommand\MeijerG*[7]{%
  G^{\,#1,#2}_{#3,#4}\MeijerM*{#5}{#6}{#7}}

\begin{document}

\title{ Mixed RF-VLC Relaying System with Radio-Access Diversity }
\author{Milica~I.~Petkovic,~\IEEEmembership{Member,~IEEE,}~Aleksandra~Cvetkovic,~\IEEEmembership{Member,~IEEE,}\\~Milan~Narandzic,~\IEEEmembership{Member,~IEEE,}~Dejan~Vukobratovic,~\IEEEmembership{Senior Member,~IEEE}
\thanks{This work has received funding from the European Union Horizon 2020 research and innovation programme under the Marie Skodowska-Curie grant agreement No 734331. 
The work  was partially supported by Ministry of Education, Science and Technology Development of Republic of Serbia under grants TR-32025 and III44003.}
\thanks{M.~I.~Petkovic, M. Narandzic and D. Vukobratovic are with University of Novi Sad, Faculty of Technical Science, Novi Sad, Serbia (e-mails: milica.petkovic@uns.ac.rs; orange@uns.ac.rs; dejanv@uns.ac.rs).}
\thanks{A.~Cvetkovic is with University of Nis, Faculty of Electronic Engineering, Nis, Serbia (e-mail: aleksandra.cvetkovic@elfak.ni.ac.rs).}
}

\maketitle

\begin{abstract}
We present a statistical analysis of a  mixed  radio-frequency (RF)-visible light communications (VLC) relaying system, where outdoor millimeter wave based RF links are utilized to provide backhaul connectivity for indoor VLC broadcasting. The multiple RF links are assumed to communicate with the VLC access point through decode-and-forward relay. Novel closed-form outage probability and average bit error rate expressions are derived and utilized to obtain numerical results. Monte Carlo simulations validate presented numerical results, which are further used to examine the effects of system and  channel parameters on system performance.
\end{abstract}


\IEEEpeerreviewmaketitle

\section{Introduction}

Wireless communication systems are challenged to
fulfill novel demanding requirements and provide wider coverage for all users, whose number is constantly increasing.
For that reason, there is a need to introduce novel techniques, and/or to combine existing ones.
With regard to radio-frequency (RF) systems, millimeter wave (mmWave)  
communications provide quite large bandwidths and high data rates, thus the use of mmWave band  in urban outdoor
environments of the 5$^{\rm th}$ generation  wireless networks is certain \cite{mmW1,mmW2}.
Due to many advantages, such as support for more users, license-free operation and large bandwidth, the optical wireless communications (OWC) systems represent an appropriate alternative or complement to the  RF signal transmission   \cite{OWC_MATLAB,bookvlc}. 
While the free-space optics (FSO) represents the line-of-sight (LoS) setup of the OWC system in outdoor enviroment, the  visible light communications (VLC)  are utilized as high data-rate indoor OWC access technology \cite{OWC_MATLAB,bookvlc, ComMag,vlc1,Haas}. 
 
As an additional way to provide wider  and energy-efficient coverage area, as well as
an increased capacity, relaying technology is adopted considering different technologies.
The outdoor FSO system was  proposed  as  the last-mile access network in mixed RF-FSO relaying system in \cite{lee}.
Further,  combination of the indoor VLC  and  high bandwidth FSO 
 was observed in \cite{exp,model}. 
While \cite{exp} provided the analysis of the  experimental demonstration of a hybrid  FSO-VLC network,  \cite{model} presented the statistical analysis of  a cascaded FSO-VLC relaying system for the first time, deriving of the  outage probability and the average bit error rate (BER) expressions.

Inspired by aforementioned studies, we analyse the mixed RF-VLC relaying system,  considering that the broadband service  is provided to the end user  by the indoor VLC access point with support of the backhaul RF links. The utilization of the RF links as backhaul ones can be beneficial as a backup of unoperational FSO link in FSO-VLC systems proposed in \cite{model}.
Considered scenario assumes that the RF part of the system represents dense deployment of mmWave fifth generation (5G) backhaul links in urban scenario  \cite{mmWBC1,mmWBC2}. 
Multiple base stations (BSs) are connected by high-capacity links and use coordinated multi-point (CoMP) transmission, 
thus can share channel state information, as well as the full data signals of the users   \cite{BS1, BS2}. 
The RF links experience Rician fading, which is proved to be a convenient statistical model for describing line-of-sight (LoS) wireless backhaul  mmWave based links \cite{makki2017}.
The BSs broadcast the information to  decode-and-forward (DF) relay collocated with  VLC access point. At the relay, maximal-ratio combining (MRC) diversity technique is employed. 
The utilization of the VLC access point in  indoor  scenario can be valuable when RF unfriendly and/or highly secured environment  is required \cite{Ray1}. The example of  benefits of  VLC  implementation in indoor environment can be found in  medicine when the impact of RF frequencies on humans is unwanted    and  should  be avoided. 
Moreover, since strong intensity of the electric field,  induced by some RF frequencies, can  interfere with electronic equipments, the  incorrect data  can occur during some measurments. 
Additionally,  implementation of the VLC system can be simultaneously used for  illumination and data communications. The VLC can be also complement link to the indoor RF links for  throughput enhancement or offloading from RF to VLC links \cite{OWC_MATLAB,bookvlc, ComMag}.

To the best of the authors' knowledge, the statistical analysis of the radio-access diversity over mixed dual-hop RF-VLC relaying system has not been investigated yet. 
The relevance of such a scenario will increase in upcoming years with the deployment of mmWave 5G small cells and proliferation of indoor VLC technologies.
The main goal of this work is to derive novel analytical expressions for the outage probability and the average BER, which are utilized to examine the impact of system parameters  on the  performance. Monte Carlo simulations are used to confirm the derived analytical
results.


\section{System and channel model}

\begin{figure}[!b]
\centering
\includegraphics[width=2.3in]{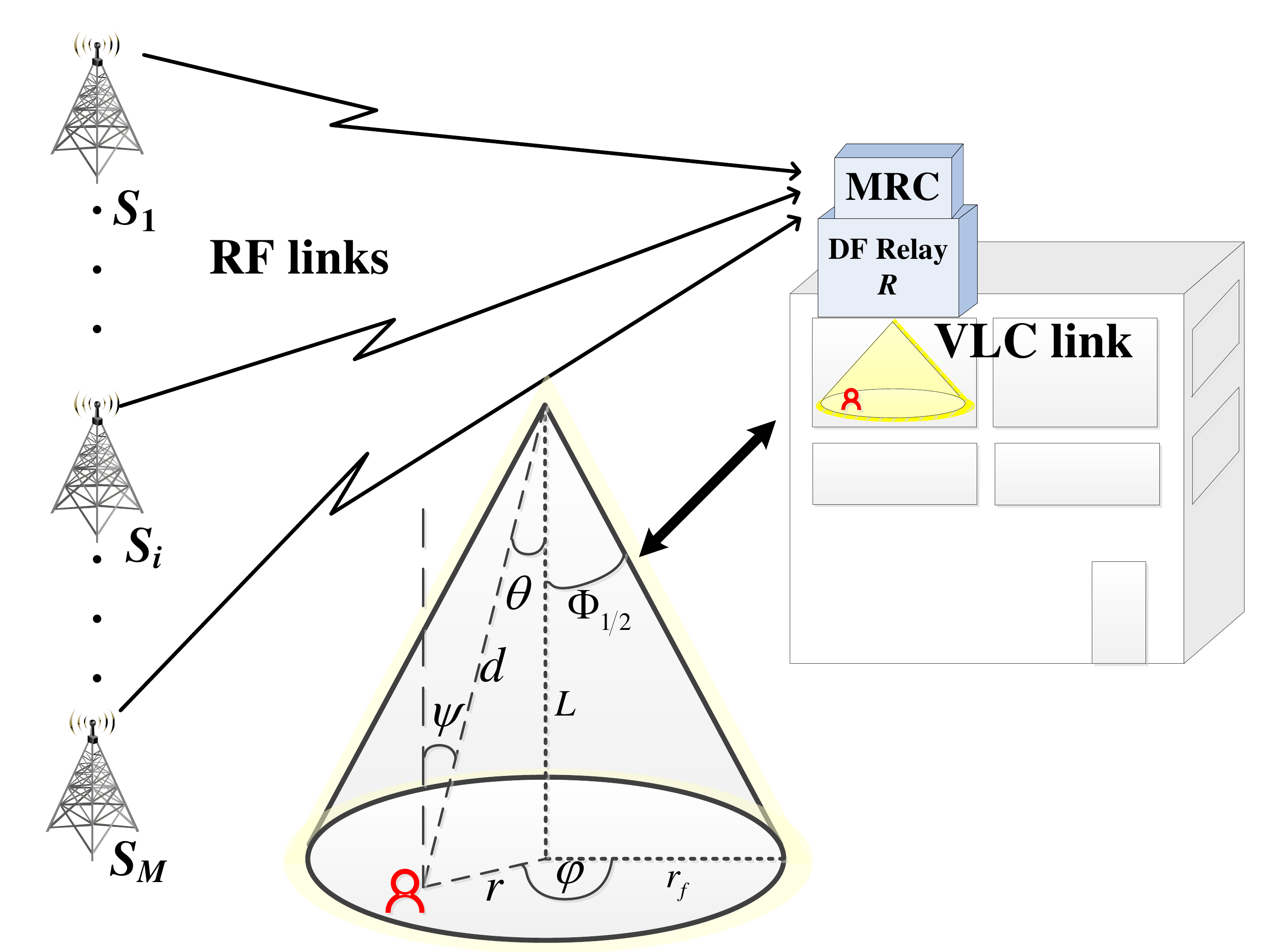}
\caption{System model of a dual-hop mixed RF-VLC communication system.}
\label{Fig1}
\end{figure}

Dual-hop mixed RF-VLC relay network is investigated. As shown in Fig.~\ref{Fig1}, considered system model includes $ M $ BSs, denoted by $ S_i $, $ i=1,\ldots,M $,  which perform transmission via RF links subject to Rician fading. Let $s_i$  denotes the signal sent from the $i$-th BS, $S_i$,  with the average transmitted electrical power  $P_s$. 
The received RF electrical signal sent from  $S_i$  is defined as
\begin{equation}
r_i = h_is_i + n_R,
\label{rR}
\end{equation}
where $h_i$  represents the  fading amplitude of the $S_i - R$ link, and  the level of an additive white Gaussian noise (AWGN) with zero mean and variance $\sigma_R^2$  over RF links is denoted by $n_R$. 
As the optimal spatial diversity technique, the  MRC is employed at a  DF relay node, denoted by $ R $.

Unlike the RF part of the system which considers outdoor environment, the second VLC link of the cascaded RF-VLC relaying system is adequate to be implemented in indoor environment. After electrical-to-optical signal conversion, relay performs retransmission on the frequency of the visible light through  downlink VLC access point implemented in some closed space (room).  The VLC access point consists of a LED lamp, which contains a group of several LEDs \cite{ISI}. The lenses are implemented to regulate
direction and focus of the LED lighting.
The VLC access point is placed on the ceiling to deliver data to the end users  uniformly distributed over the coverage area of  room. 
At the destination of the mobile end user terminal, direct detection is done and optical-to-electrical signal  conversion is performed via PIN photodetector. The electrical signal at the  destination of a mobile end user is given by 
\begin{equation}
r_D = P_tI\eta r_R + n_D,
\label{rD}
\end{equation}
where  $r_R$ is the signal at the MRC output, $I$  is the  direct current (DC) channel gain of the LoS link between LED lamp and the end user, $\eta$ is the electrical-to-optical conversion efficiency, and $P_t$ represents the average transmitted optical power of a LED lamp \cite{OWC_MATLAB}. It is assumed that $P_t = N P_l$, where $N$ is the number of LEDs with the same power $P_l$. 
The AWGN over VLC link with zero mean and variance $\sigma_D^2$  is denoted by $n_D$, 
where $\sigma _D^2=N_0B$ with the noise spectral density $N_0$, and the baseband modulation bandwidth $B$.

\subsection{RF channel}
The instantaneous signal-to-noise ratio (SNR) at the relay (in the $i$-th RF link) is defined as $\gamma_{i} = |h_i|^2P_s/\sigma _R^2$.
The fading over $S_i - R$ channel follows Rician distribution, hence, the probability density   function (PDF) of the instantaneous SNR of the $ i $-th RF link is 
\begin{equation}
f_{\gamma_{i}}\left( \gamma  \right) = \frac{\left( K_i +1 \right)e^{-K_i}}{\mu_i}  e^{ - \frac{\left( K_i +1 \right) \gamma }{\mu_i}} I_0 \!\left( 2 \sqrt{ \frac{K_i\left( K_i +1 \right) \gamma}{\mu_i}} \right),
\label{pdfi}
\end{equation}
where $K_i$  is the Rician  factor in the $i$ -th RF link, representing the ratio of the power of the LoS path component to the average power of the scattered path component, $I_\nu (\cdot)$
represents the $\nu$-th order modified Bessel function of the first kind \cite[(8.43)]{grad}, and  $\mu_i$ denotes the average SNR  defined as statistical average of the channel power gain  $\gamma_i$, i.e., $\mu_i = {\rm E} \left[ \gamma_i \right]$ (${\rm E}\left[ \cdot \right]$ represents the mathematical expectations). Since MRC combining technique is applied, the combined electrical signal which arrives at the relay is defined as 
\begin{equation}
\gamma_{\rm rf} = \sum_{i=1}^{M} \gamma_{i}.
\label{MRC}
\end{equation}
Under the assumption of independent and identically distributed fading  RF channels, i.e., $K_{i}=K$ and $\mu_{i}=\mu_{\rm rf}$ for all $i=1,\ldots,M $, based on (\ref{pdfi}) and (\ref{MRC}), the PDF of the MRC signal at the relay is defined as \cite{rice}
\begin{equation}
\begin{split}
f_{\gamma_{\rm rf}}\left( \gamma  \right)& = \frac{\left( K +1 \right)e^{-KM}}{\mu_{\rm rf}} \left(  \frac{\left( K +1 \right)\gamma  }{K M \mu_{\rm rf}} \right)^{\frac{M-1}{2}} \\
& \times e^{ - \frac{\left( K +1 \right) \gamma }{\mu_{\rm rf}}} I_{M-1} \left( 2 \sqrt{ \frac{K\left( K +1 \right) M\gamma}{\mu_{\rm rf}}} \right),
\label{pdfRF}
\end{split}
\end{equation}
The cumulative distribution function (CDF) of the signal $\gamma_{\rm rf}$ at the relay is 
\begin{equation}
\begin{split}
F_{\gamma_{\rm rf}}\left( \gamma  \right)& = 1-  \left(  K M  \right)^{1-M} \\
& \times {\rm Q}_{M} \left(  \sqrt{2KM}, \sqrt{\frac{2\left( K +1 \right) \gamma}{\mu_{\rm rf}}}\right),
\label{cdfRF}
\end{split}
\end{equation}
where ${\rm Q}_M (a,b) = \int\limits_b^\infty  {x{{\left( {\frac{x}{a}} \right)}^{M - 1}}{e^{ - \frac{{{x^2} + {a^2}}}{2}}}{I_{M - 1}}\left( {ax} \right)} {\rm d} x$
represents the $M$-th order Marcum Q-function.

\vspace{-\baselineskip}
\subsection{VLC channel}
The second hop represents the indoor environment transmission at the frequencies of the visible light spectrum. The VLC channels include both LoS and diffuse components, but, as it was concluded in \cite{LOS}, the  energy of the LoS component  is greater than the reflected signals energy. Based on this ascertainment, the energy of the reflected signals is neglected.

The optical signal transmission is performed from the LED lamp placed at the ceiling to the single mobile end user terminal. As it is depicted in Fig.~\ref{Fig1}, the LED lamp transmitter is positioned at height $ L $ from the end user set with angle of irradiance $\theta$, and the angle $\varphi$  and radius $r$  in the polar coordinate plane. Furthermore, the angle of incidence is denoted by  $\psi$, while $ d $ is the Euclidean distance between the  LED lamp and the photodetector receiver.  Direction and focus of  transmitted optical signal is regulated by lenses implemented as a part of the LED lamp.
The LED transmitter is assumed to be modeled by a generalized Lambertian emission pattern, with the order related to the semi-angle at the half power of LED, denoted by  $\Phi_{1/2}$,  as $m =  - \ln 2/\ln \left( \cos \Phi_{1/2} \right)$ \cite{OWC_MATLAB}.
The  semi-angle at the half power of LED is related by the maximum radius of a LED cell footprint, $r_f$, as $r_f = L\sin \left(\Phi_{1/2} \right)/\cos \left( \Phi_{1/2} \right)$. Furthermore, a photodetector receiver is characterized by physical surface area  denoted by $ A $, the responsivity $\Re$, the gain of the optical filter  $T$, and the optical concentrator $g\left( \psi \right) = \rho^2/\sin^2\left( \Psi \right)$ for $0\leq\psi \leq \Psi$, where $\rho$ is the refractive index of lens at a photodetector, and $\Psi$ denotes the field of view (FOV) of the receiver.

Based on aforementioned definitions, the DC channel gain of the LoS link between LED and the mobile end user receiver at the distance $ d $ and angle  $ \theta $ with respect to transmitter, is \cite{OWC_MATLAB}
\begin{equation}
I = \frac{A\left( m + 1\right)\Re}{2\pi d^2}\cos ^m\left( \theta \right)Tg\left( \psi \right)\cos \left( \psi \right).
\label{I_n1}
\end{equation}
The assumption that the surface of photodetector receiver is parallel to the ground plane and has no orientation towards the LED was taken into account, i.e., $\theta =\psi $. From Fig.~\ref{Fig1}, it can be concluded that $d~=~\sqrt {r^2 + L^2} $  and  $\cos \left( \theta \right) \!=\! \frac{L}{ \sqrt {r^2 + L^2} }$.  
After replacement, (\ref{I_n1}) is rewritten as $I = \Upsilon /\left( r^2 + L^2 \right)^{\frac{m + 3}{2}} $, where  $\Upsilon  = \frac{A\left( m + 1 \right)\Re}{2\pi}Tg\left( \psi \right)L^{m + 1}$.
 Furthermore, it is assumed that position of mobile end user is random over
circular area covered by LED lighting, being  modeled by a uniform distribution, thus  the probability density function (PDF) of the  radial distance is $f_{r}\left( r \right) = 2r / r_f^2,$ $0\leq r \leq r_f $.
After applying technique for transformation of random variables, the PDF of the channel gain, $ I $, is derived as 
\begin{equation}
f_{I}\left( I \right) = \frac{2}{r_f^2\left( m + 3 \right)} \Upsilon ^{\frac{2}{m + 3}}I^{ - \frac{m + 5}{m + 3}},\quad I_{\min }\leq I \leq I_{\max },
\label{pdf_In}
\end{equation}
where $I_{\min} = \frac{\Upsilon }{{\left( r_f^2 + L^2 \right)}^{\frac{m + 3}{2}}}$ and  $I_{\max} = \frac{\Upsilon}{L^{m + 3}}$.

The instantaneous SNR of the VLC channel of the  end user is defined as 
\begin{equation}
\gamma_{\rm vlc} = \frac{P_t^2\eta ^2I^2}{\sigma _D^2}.
\label{snrVLC}
\end{equation}
Based on (\ref{pdf_In}) and (\ref{snrVLC}), the PDF of the instantaneous SNR of the  end user is derived as
\begin{equation}
f_{\gamma_{\rm vlc}}\left( \gamma  \right) = \frac{\mu_{\rm vlc}^{\frac{1}{m + 3}} \Upsilon^{\frac{2}{m + 3}}}{r_f^2\left( m + 3 \right)}\gamma^{ - \frac{m + 4}{m + 3}},\quad \gamma_{\min }\leq \gamma \leq \gamma_{\max },
\label{pdfVLC}
\end{equation}
where $\gamma_{\min} = \frac{\mu_{\rm vlc}  \Upsilon^2}{{\left( r_f^2 + L^2 \right)}^{m + 3}}$ and  $\gamma_{\max} = \frac{\mu_{\rm vlc} \Upsilon^2}{L^{2 \left(m + 3\right)}}$, and $\mu_{\rm vlc}~=~\frac{P_t^2\eta ^2}{\sigma _D^2}$. Furthermore, after performing integration, the CDF of the instantaneous SNR of the  end user is derived as
\begin{equation}
F_{\gamma_{\rm vlc}}\!\!\left( \gamma  \right)\!\! =\!\! \left\{ {\begin{array}{*{20}{c}}
\!\!\!{1\! +\! \frac{{{L^2}}}{{r_f^2}}\! - \!\frac{{{\Upsilon ^{\frac{2}{{m + 3}}}}}}{{r_f^2}}{{\left( {\frac{\gamma }{{{\mu _2}}}} \right)}^{ \!\!-\! \frac{1}{{m + 3}}}},}  \gamma_{\min}\!\!\leq \!\!\gamma \leq\gamma_{\max}\\
{1,}  \quad \quad \quad \quad \quad \quad \quad \quad \quad \quad \gamma > \gamma_{\max}\\
\end{array}} \right.\!\!\!\!.
\label{cdfVLC}
\end{equation}

\vspace{-\baselineskip}
\subsection{Statistic of the equivalent end-to-end SNR}

For considered DF based RF-VLC relaying system, the instantaneous equivalent end-to-end SNR, $ \gamma_{\rm eq} $, of the mobile end user  is defined as $\gamma_{\rm eq} = \min \left( \gamma _{\rm rf},\gamma _{\rm vlc} \right)$,
where $\gamma _{\rm rf}$  and $\gamma _{\rm vlc}$   are the instantaneous SNRs of the RF and the  VLC links, respectively.
The CDF of $ \gamma_{\rm eq} $ can be determined as 
\begin{equation}
\begin{split}
&F_{\gamma_{\rm eq}}\left( \gamma \right) = \Pr \left[ \gamma_{\rm eq} < \gamma \right] \\
& = F_{\gamma _{\rm rf}}\left( \gamma \right) + F_{\gamma _{\rm vlc}}\left( \gamma \right) - F_{\gamma _{\rm rf}}\left( \gamma \right)F_{\gamma _{\rm vlc}}\left( \gamma  \right),
\end{split}
\label{Pout}
\end{equation}
where  the CDFs   $F_{\gamma _{\rm rf}}\left(  \cdot  \right)$ and  $F_{\gamma _{\rm vlc}}\left(  \cdot  \right)$  are previously defined in (\ref{cdfRF}) and (\ref{cdfVLC}), respectively.


\section{Mixed RF-VLC system  performance}


The outage probability defines the probability that the instantaneous equivalent end-to-end SNR falls below a predetermined outage protection value, $ \gamma_{\rm th} $. 
 The outage probability of the mixed RF-VLC system for a single end user can be obtained as $P_{out} = F_{\gamma_{\rm eq}}\left( \gamma_{\rm th} \right) $ based on (\ref{Pout}).


 The average BER expression for DF based  RF-VLC relaying system is 
\begin{equation}
P_e = P_{e_{\rm rf}}\left( 1 - P_{e_{\rm vlc}} \right) + P_{e_{\rm vlc}}\left( {1 - P_{e_{\rm rf}}} \right),
\label{Pe2}
\end{equation}
where $P_{e_{\rm rf}}$ and $P_{e_{\rm vlc}}$  represent the average BER of the RF and VLC links, respectively. If binary phase-shift keying  is employed, the average BER of the RF and VLC links can be  determined as \cite{proakis}
\begin{equation}
P_{e_{\rm rf}}=  \frac{1}{2}\int\limits_{0 }^{\infty } {\rm erfc}\left(  \sqrt{ \gamma} \right) f_{\gamma _{\rm rf}} \left( \gamma  \right){\rm d} \gamma ,
\label{intRF}
\end{equation}
\vspace{-\baselineskip}
\begin{equation}
~P_{e_{\rm vlc}}=  \frac{1}{2}~\int\limits_{0 }^{\infty } {\rm erfc} \left( \sqrt{ \gamma} \right) f_{\gamma _{\rm vlc}} \left( \gamma  \right){\rm d} \gamma,
\label{intVLC}
\end{equation}
where  the PDFs  $f_{\gamma _{\rm rf}}\left( \gamma  \right)$ and $f_{\gamma _{\rm vlc}}\left( \gamma  \right)$  are defined in (\ref{pdfRF}) and (\ref{pdfVLC}), respectively, and ${\rm erfc}\left( \cdot\right)$ represents the complementary error function defined in \cite[(8.250.4)]{grad}.

Firstly, we consider the RF link. 
After substituting  (\ref{pdfRF}) into (\ref{intRF}), the average BER  of the RF link is re-written as 
\begin{equation}
\begin{split}
& P_{e_{\rm rf}}=  \frac{\left( K +1 \right)e^{-KM}}{2\mu_{\rm rf}} \left(  \frac{ K +1 }{K M \mu_{\rm rf}} \right)^{\frac{M-1}{2}}  \int\limits_{0 }^{\infty }\gamma^{\frac{M-1}{2}} \\
&\!\times  e^{ - \frac{\left( K +1 \right) \gamma }{\mu_{\rm rf}}} {\rm erfc} \left(  \sqrt{ \gamma} \right)   I_{M-1} \!\left(\! 2 \sqrt{ \frac{K\left( K +1 \right) M\gamma}{\mu_{\rm rf}}} \right){\rm d} \gamma,
\label{intRF1}
\end{split}
\end{equation} 
In order to solve integral in (\ref{intRF1}), \cite[(03.02.06.0037.01)]{sajt} is utilized 
to present the modified Bessel function of the first kind into a series form as
\begin{equation}
\begin{split}
&I_{M-1} \left( 2 \sqrt{ \frac{K\left( K +1 \right) M\gamma}{\mu_{\rm rf}}} \right) =\\
&=  \sum_{k=0}^{\infty} \frac{1}{k!\Gamma\left( M +k \right)}  \left(\frac{K\left( K +1 \right) M\gamma}{\mu_{\rm rf}}\right)^{\frac{M-1}{2}+k},
\label{Bessel}
\end{split}
\end{equation}
After substituting (\ref{Bessel}) into (\ref{intRF1}), integral in (\ref{intRF1}) is 
\begin{equation}
\begin{split}
P_{e_{\rm rf}}&=  \frac{\left( K +1 \right)e^{-KM}}{2\mu_{\rm rf}} \left(  \frac{\ K +1 }{K M \mu_{\rm rf}} \right)^{\frac{M-1}{2}} \\
& \times \sum_{k=0}^{\infty} \frac{1}{k!\Gamma\left( M +k\right)}  \left(\frac{K\left( K +1 \right) M}{\mu_{\rm rf}}\right)^{\frac{M-1}{2}+k} \\
& \times \int\limits_{0 }^{\infty }\gamma^{M+k-1}  e^{ - \frac{\left( K +1 \right) \gamma }{\mu_{\rm rf}}}{\rm erfc}\left(  \sqrt{ \gamma} \right) {\rm d} \gamma.
\label{intRF2}
\end{split}
\end{equation}
After applying \cite[(01.03.26.0004.01)]{sajt} to represent the exponential function in terms of the Meijer's \textit{G}-function  as $e^{ - \frac{\left( K +1 \right) \gamma }{\mu_{\rm rf}}} =  \MeijerG*{1}{0}{0}{1}{ - }{0}{ \frac{\left( K +1 \right) \gamma }{\mu_{\rm rf}}}$ 
and \cite[(06.27.26.0006.01)]{sajt} to represent the complementary error function in terms of the Meijer's \textit{G}-function as ${\rm erfc}\left(  \sqrt{ \gamma} \right) = \frac{1}{\sqrt{\pi}} \MeijerG*{2}{0}{1}{2}{ 1 }{0,\, \frac{1}{2}}{ \gamma},$
the average BER in (\ref{intRF2}) is re-written as
\begin{equation}
\begin{split}
&P_{e_{\rm rf}}=  \frac{\left( K +1 \right)e^{-KM}}{2\mu_{\rm rf}\sqrt{\pi}} \left(  \frac{ K +1  }{K M \mu_{\rm rf}} \right)^{\frac{M-1}{2}} \\
& \times \sum_{k=0}^{\infty} \frac{1}{k!\Gamma\left( M+k \right)}  \left(\frac{K\left( K +1 \right) M}{\mu_{\rm rf}}\right)^{\frac{M-1}{2}+k} \\
&\! \!\!\times \int\limits_{0 }^{\infty }\gamma^{M+k-1}   \MeijerG*{1}{0}{0}{1}{ - }{0}{ \frac{\left( K +1 \right) \gamma }{\mu_{\rm rf}}}\MeijerG*{2}{0}{1}{2}{ 1 }{0,\, \frac{1}{2}}{ \gamma} {\rm d} \gamma.
\label{intRF3}
\end{split}
\end{equation}
In order to  solve integral in (\ref{intRF3}), \cite[(07.34.21.0011.01)]{sajt} is used. The  expression for the average BER $P_{e_{\rm rf}}$ is derived as
\begin{equation}
\begin{split}
P_{e_{\rm rf}}&=  \frac{e^{-KM}}{2\sqrt{\pi}}  \sum_{k=0}^{\infty} \frac{ \left(K M\right)^{k} }{k!\Gamma\left( M +k \right)} \MeijerG*{2}{1}{2}{2}{ 1-k-M,\,1 }{0,\, \frac{1}{2}}{\frac{\mu_{\rm rf} }{K +1}}.
\label{PeRF}
\end{split}
\end{equation}

The average BER of the VLC link is obtained after substituting (\ref{pdfVLC}) into (\ref{intVLC}), as
\begin{equation}
\begin{split}
P_{e_{\rm vlc}} = \frac{\mu_{\rm vlc}^{\frac{1}{m + 3}} \Upsilon^{\frac{2}{m + 3}}}{2  r_f^2\left( m + 3 \right)}\int\limits_{\gamma _{\min }}^{\gamma _{\max }}\!\!\! \gamma ^{ - \frac{m + 4}{m + 3}}{\rm erfc} \left( \sqrt \gamma  \right) {\rm d} \gamma
\end{split}.
\label{intVLC1}
\end{equation}
Integral in (\ref{intVLC1}) is solved by \cite[(06.27.21.0005.01)]{sajt}, and the closed-form expression for $P_{e_{\rm vlc}}$ is derived  as
\begin{equation}
\begin{split}
&P_{e_{\rm vlc}} = \frac{\mu_{\rm vlc}^{\frac{1}{m + 3}} \Upsilon^{\frac{2}{m + 3}}}{2r_f^2}\\
&  \times \left( \frac{1}{\sqrt{\pi}} \Gamma \left( \frac{m+1}{2m + 6},\gamma_{\max} \right) - \gamma_{\max}^{-\frac{1}{m+3}} {\rm erfc}\left( \sqrt{ \gamma_{\max} } \right)  \right. \\
& \left.-\frac{1}{\sqrt{\pi}}\Gamma \left( \frac{m+1}{2m + 6},\gamma_{\min} \right)  + \gamma_{\min}^{-\frac{1}{m+3}} {\rm erfc}\left( \sqrt {\gamma_{\min }} \right) \right), 
\end{split}
\label{PeVLC}
\end{equation}
where $\Gamma \left(  \cdot, \cdot \right)$ is Incomplete Gamma function  \cite[(8.350)]{grad}.

After substituting (\ref{PeRF}) and (\ref{PeVLC}) into (\ref{Pe2}), the closed-form expression for the average BER is derived.

\begin{figure}[!t]
\centering
\includegraphics[width=2.8in]{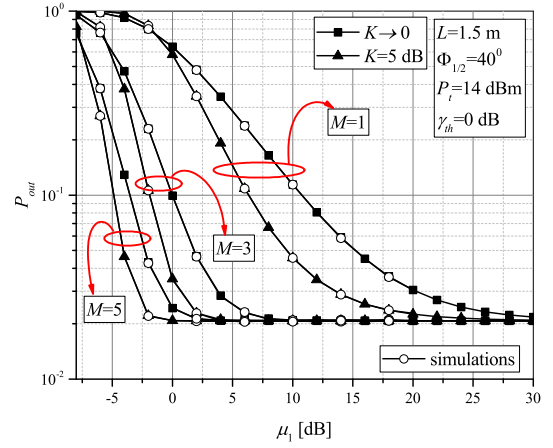}
\caption{Outage probability vs. average SNR over RF channel.}
\label{Fig2}
\end{figure}

\section{Numerical results and discussions}

Based on derived expressions for the outage probability and the average BER, numerical results are presented. Monte Carlo simulations are utilized to  validate derived expressions. Following values of the parameters are assumed:  the FOV of the receiver $\Psi=60^0$, the photodetector surface area $A=1~{\rm cm}^2$, the responsivity $\Re=0.4~{\rm A}/{\rm W}$, the optical filter  gain $T =1$, the refractive index $\varrho=1.5$. Furthermore, the  electrical-to-optical conversion efficiency is $\eta=0.8$, noise spectral density takes a value $N_0=10^{-21}~{\rm W}/{\rm Hz}$, and the baseband modulation bandwidth is $B= 20~{\rm MHz}$ \cite{vlc1, Haas}.

Fig.~\ref{Fig2}  depicts the outage probability of the  RF-VLC system  dependence on the average SNR over RF link, considering different numbers of BSs. 
Significant improvement of the overall system
performance with greater $M$ can observed.
Furthermore, the impact of the Rician factor is considered. System performs better when the Rician factor is greater, i.e., $K=5$ dB than when it tends to zero $K\rightarrow 0$, which corresponds to the Rayleigh  fading distributed RF links. It can be observed that the effect of the Rician fading strength of the RF links has more impact on the system performance when lower number of BSs is assumed. With utilization of the greater number of BSs when MRC is employed, the effect of Rician fading can be considerably reduced.
In addition, a certain outage probability floor is noticed, meaning that further increase of the RF signal power will not lead to the system performance improvement. This outage floor  appears at lower values of $\mu_1$ when greater number of BSs is present and when the Rician factor is higher.

\begin{figure}[!t]
\centering
\includegraphics[width=2.8in]{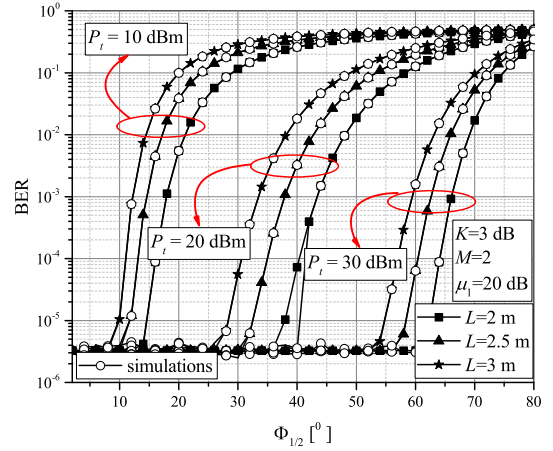}
\caption{Average BER vs. the semi-angle at the half power of LED.}
\label{Fig3}
\end{figure}

In  Fig.~\ref{Fig3},  the average BER dependence on the semi-angle at the half power of LED lamp is shown. 
 With greater $\Phi_{1/2}$, the received optical power is reduced due to large amount of dissipation i.e., distribution of energy over an excessively large area.
 On the other hand, when $\Phi_{1/2}$ is smaller, the optical signal is narrower and more focused, thus the received optical power is greater and system has better  performance. In addition,  the impact of $\Phi_{1/2}$
  on the average BER impairment is dependent on $P_t$: for lower $P_t$ BER impairment becomes significant for lower $\Phi_{1/2}$ values.
  Results illustrate that  the  performance deteriorates with greater room height $L$. Since the optical signal propagation path is longer, the total received power will be reduced. 
 Furthermore, it is noted that the room height has no influence on the BER performance for lower semi-angles, especially when the LED optical power is greater. This is valid due to the fact that the small $\Phi_{1/2}$ corresponds to very narrow and directed  optical signal. Thus, there will be no dissipation of the optical energy, and the propagation path loss has a minor influence on the transmission quality.

\begin{figure}[!t]
\centering
\includegraphics[width=2.8in]{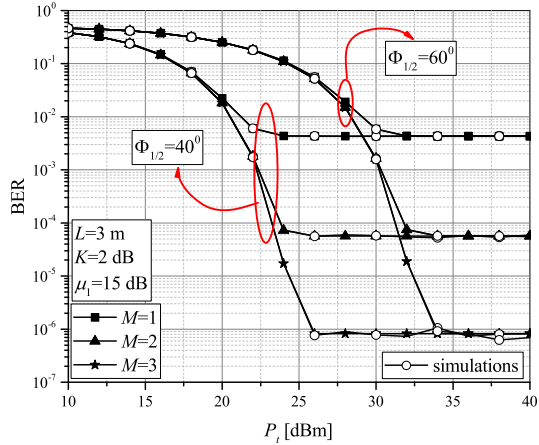}
\caption{Average BER  vs. transmitted optical power.}
\label{Fig4}
\end{figure}

The average BER vs.  transmitted optical power is presented in Fig.~\ref{Fig4}. 
Impact of   number of  BSs on BER is stronger when $\Phi_{1/2}$ is smaller in the range of small optical power. 
A certain average BER floor is observed  when  the  RF signal power  is constant, thus further increase of the optical power will not result in better system performance. 
It is dependent  on the number of RF links, but independent on the semi-angle at the half power of LED lamp. Analogously, it can be concluded that the value of the BER floor, appearing for certain values of the optical power,  depends on the RF link parameters, but it is independent on VLC access point specification. 
 Furthermore,  the BER floor appears at lower values of $P_t$ when is $\Phi_{1/2}$ smaller, as well as for lower number  $M$. This average BER floor, as well as the outage floor, appears as limiting factor that should be managed in RF-VLC system design. 

\section{Conclusion}
In this paper, we have presented the statistical analysis of the  RF-VLC relaying system.  It has been concluded that the semi-angle at the half power of LED significantly affects the RF-VLC system performance. Lower value of the angle results in narrower optical signal, which leads to greater received optical power.
Lower room, related to the VLC  environment, reflects in better system performance, since the propagation path of the optical signal transmission is shorter. 
Furthermore,  the certain  outage and average BER floors are noticed,  causing that  further increase in electrical or optical power will not lead to the system performance improvement, which  should be considered in RF-VLC system design.

\end{document}